\documentclass[11pt,a4paper]{article}
\usepackage[T1]{fontenc}
\usepackage[utf8]{inputenc}
\usepackage{lmodern,microtype}
\usepackage{setspace}
\usepackage[margin=28mm]{geometry}
\usepackage{amsmath,amssymb,amsthm,mathtools}
\usepackage{booktabs,array,float}
\usepackage{graphicx}
\usepackage{placeins}
\usepackage[authoryear,round,sort]{natbib}
\usepackage{multibib}
\newcites{app}{Online Appendix References}
\usepackage[colorlinks=true,linkcolor=blue,citecolor=blue,urlcolor=blue]{hyperref}
\usepackage[nameinlink,noabbrev]{cleveref}
\allowdisplaybreaks
\newtheorem{theorem}{Theorem}[section]
\newtheorem{proposition}[theorem]{Proposition}
\newtheorem{lemma}[theorem]{Lemma}

\theoremstyle{definition}

\theoremstyle{remark}

\DeclareMathOperator{\Rea}{Re}
\DeclareMathOperator{\Ima}{Im}
\DeclareMathOperator{\Log}{Log}
\DeclareMathOperator{\sgn}{sgn}

\newcommand{\ii}{\mathrm{i}}
\newcommand{\dd}{\,\mathrm{d}}
\newcommand{\R}{\mathbb R}
\newcommand{\C}{\mathbb C}
\newcommand{\E}{\mathbb E}

\title{A general technique for explicit quantile representations}
\author{Wolfgang Schadner\thanks{University of Liechtenstein, wolfgang.schadner@uni.li}}
\date{\today}

\begin{document}
\maketitle
\begin{abstract}
	This work proves existence of contours independent of probability level for exact quantile integrals of strictly positive real-analytic densities. Explicit constructions cover five broad distribution families and yield a finite-interval representation of Black--Scholes implied volatility using forward functions.
\end{abstract}

\section{Introduction}\label{sec:introduction}
An explicit distribution function need not have a familiar closed-form inverse. I seek exact quantile representations using forward functions and a contour chosen without knowing the quantile. For a working distribution with CDF $\mathcal P$ and quantile $\mathcal Q=\mathcal P^{-1}$, one useful form is
\begin{equation}\label{eq:intro}
\mathcal Q(p)=\frac1\pi\Ima\int_I
\kappa'(x)\Log\left(\frac{\mathcal P(\kappa(x))-p}{1-p}\right)\dd x,
\qquad 0<p<1.
\end{equation}
Here $I$ is the working support interval, and the contour above it has the form
\begin{equation}\label{eq:contour}
\kappa(x)=x+\ii h(x),\qquad x\in I.
\end{equation}
Constructing the contour height $h$ from the density, independently of the probability level $p$, is the main obstacle addressed here. The contour must isolate the real quantile and ensure convergence of the integral. Choosing its height independently of $p$ allows the same contour to serve every probability level.

Equation~\eqref{eq:intro} follows by integration by parts from the classical logarithmic-derivative formula for an isolated zero \citep{DelvesLyness1967}. Riemann boundary-value methods likewise yield explicit expressions for roots and inverse functions \citep{BurnistonSiewert1973,PickettMillev2002}. For the contour identity used here, the essential construction problem is to isolate the required root and control the endpoint contributions. Other approaches characterize quantiles through differential equations and power series \citep{SteinbrecherShaw2008}, construct Taylor approximations from density derivatives \citep{YuZelterman2017}, or invert saddlepoint approximations to obtain approximate quantiles \citep{Arevalillo2003}. I address the contour construction directly. Density bounds determine a path independently of the unknown quantile, yielding an exact integral of forward functions valid for every probability level. The general existence theorem and the elementary heights derived for selected families make this construction applicable beyond a single inversion problem. \\

The construction starts from the positive density on the real axis. Choose a height at each point so that the density's real part stays positive along the vertical segment. The complex CDF then has positive imaginary part above the axis and cannot equal a real probability there. Endpoint estimates complete the argument. For strictly positive real-analytic densities on an interval, I prove existence after a known coordinate change, without moment assumptions. I also give a quantitative prescription when complex-disk bounds are available and derive elementary heights for selected families. \\

Section~\ref{sec:method} develops the construction using generalized inverse Gaussian (GIG) as a working example. Section~\ref{sec:families} gives contours and reconstruction maps for generalized beta, generalized gamma, generalized hyperbolic, and generalized extreme-value laws, often sharing one working density across several subfamilies.  Section~\ref{sec:applications} illustrates the method's reach with a finite-interval representation of Black--Scholes implied volatility involving only elementary functions and the error function. The appendix supplies proofs, exact sanity checks, and numerical verifications, together with an extension to decreasing analytic functions and a representation of Bachelier implied volatility.

\section{Method}\label{sec:method}\label{sec:general}
Let $g=\mathcal P'>0$ be a normalized working density on an open interval $I$. For a positive, locally Lipschitz height $h$, define the ribbon $\mathcal R_h=\{x+\ii y:x\in I,\ |y|\le h(x)\}$ between $\kappa$ and its conjugate. Assume that $\mathcal P$ and $g$ extend holomorphically to a neighborhood of this ribbon and that $\Ima\mathcal P(x+\ii y)>0$ for $0<y\le h(x)$. This sign condition and conjugation exclude nonreal zeros of $\mathcal P-p$. The unique real zero is $\mathcal Q(p)$, and it is simple because $g>0$.

On the positively oriented boundary $\mathcal C$ of a truncated ribbon containing this zero, the classical root-moment identity gives \citep{DelvesLyness1967}
\begin{equation}\label{eq:quantile-root}
\frac{1}{2\pi\ii}\oint_{\mathcal C} z\frac{g(z)}{\mathcal P(z)-p}\dd z=\mathcal Q(p).
\end{equation}
The upper arc runs right to left, so the two conjugate arcs contribute $-2\ii$ times the imaginary part of the upper integral taken left to right. If that integral converges absolutely and the vertical closing integrals vanish at the support endpoints, then
\begin{equation}\label{eq:quantile-density}
\mathcal Q(p)=-\frac1\pi\Ima\int_I
\kappa(x)\kappa'(x)\frac{g(\kappa(x))}{\mathcal P(\kappa(x))-p}\dd x,
\qquad 0<p<1.
\end{equation}
Here $\kappa'$ exists almost everywhere. Appendix~\ref{app:proofs} collects the proofs, endpoint criteria, and extensions used in this section.

\begin{theorem}[Existence of a contour independent of the probability level]\label{thm:universal}
Every strictly positive real-analytic normalized density on an open interval admits an increasing real-analytic change of coordinate to $(0,1)$ and a positive Lipschitz height in that coordinate satisfying the preceding inversion conditions for all $0<p<1$. The height is independent of $p$, and the known coordinate map recovers the original quantile. No moment assumption is required.
\end{theorem}
The theorem guarantees existence, with a quantitative construction available when suitable complex-disk bounds are known. Obtaining an elementary height requires further estimates specific to the density, as supplied for the examples below. Here explicit quantile representations mean exact integrals involving forward functions along specified contours. The following five steps develop this construction, with GIG illustrating each step.

\subsection{Step 1: Density}\label{sec:step-density}
Choose a normalized working density $g>0$ on its open support interval $I$, real analytic in the interior. Endpoint singularities are allowed. A known real-analytic monotone map $u=T(x)$ with nonzero derivative can simplify the density. The working density is then the original density evaluated at $T(x)$, multiplied by $|T'(x)|$. Retain the map and its direction for reconstruction in Step~5. I use $g$ and $\mathcal P$ for the working density and CDF, and $f$ for the standardized family PDF.

If the density is not analytic at an interior point, work separately on its analytic intervals and normalize their probability masses. The CDF values at the dividing points determine the relevant interval and adjusted probability level. Each interval requires its own contour and endpoint checks.

\paragraph*{Example.}
For the GIG law, take $x=u/\sigma$ and use the standardized density \citep{Jorgensen1982}
	\begin{equation}\label{eq:gigdensity}
	I=(0,\infty),\qquad
	g(x)=f(x)=\frac{x^{a-1}}{2K_a(b)}
	\exp\!\left[-\frac b2\left(x+\frac1x\right)\right].
	\end{equation}
Here $a\in\R$ is the order, $b>0$ the concentration, $\sigma>0$ the scale, and $K_a$ the modified Bessel function of the second kind. Write $\mathcal P_{\mathrm{GIG}}$ and $\mathcal Q_{\mathrm{GIG}}$ for the standardized CDF and quantile. Continuing powers from the positive real axis makes $g$ holomorphic on $\Rea z>0$.

\subsection{Step 2: Bound}\label{sec:step-bound}
Write $z=x+\ii y$ and continue $g$ holomorphically to a simply connected, symmetric neighborhood of $I$, with branches fixed by the real density. Normalize its primitive to agree with $\mathcal P$ on $I$. Positive real part along each vertical segment ensures
	\begin{equation}\label{eq:vertical}
	\Rea g(x+\ii t)>0\quad(0\le t\le y)
	\quad\Longrightarrow\quad
	\Ima\mathcal P(x+\ii y)=\int_0^y\Rea g(x+\ii t)\dd t>0.
	\end{equation}
Two types of estimate give this positivity. A relative variation bound
	\begin{equation}\label{eq:relative}
	|g(x+\ii y)-g(x)|\le\delta g(x),\qquad 0<\delta<1,
	\end{equation}
implies $\Rea g(x+\ii y)\ge(1-\delta)g(x)>0$ and controls the modulus. Alternatively, where $g$ admits a holomorphic logarithm real on $I$, the phase bound
	\begin{equation}\label{eq:phase}
	|\arg g(x+\ii y)|
	\le\int_0^y\left|\frac{g'(x+\ii t)}{g(x+\ii t)}\right|\dd t
	<\frac{\pi}{2}
	\end{equation}
also ensures $\Rea g>0$. Evaluating the phase directly can sharpen this bound. Here $\arg g$ denotes the continuous argument normalized to zero on $I$. On the verified ribbon it agrees with the principal argument. Both estimates must hold throughout the vertical segment. I use relative variation for generalized hyperbolic (GH) and variance-gamma (VG) laws, and phase bounds for beta, gamma, and GIG.
	
	\paragraph*{Example.}
	For GIG, the continuous phase is
	\begin{equation}
	\arg g(x+\ii y)
	=(a-1)\arctan\frac yx-\frac{by}{2}
	+\frac{by}{2(x^2+y^2)}.
	\end{equation}
For $y\ge0$, the triangle inequality and $\arctan(y/x)\le y/x$ therefore give
	\begin{equation}
	|\arg g(x+\ii y)|
	\le y\left(\frac{|a-1|}{x}+\frac b2+\frac{b}{2x^2}\right).
	\end{equation}

\subsection{Step 3: Height}\label{sec:step-height}
Let $h_0(x)>0$ be a height up to which the vertical segment remains in the holomorphic domain and the estimate from Step~2 is valid. Choose a positive, locally Lipschitz height $h(x)\le h_0(x)$, independently of $p$, so that throughout $0\le y\le h(x)$, the phase bound is strictly below $\pi/2$, or the relative variation bound is strictly below one. Either condition ensures $\Rea g(x+\ii y)>0$.

In particular, if either estimate is bounded by $y/(2h_1(x))$ for $0\le y\le h_0(x)$, a sufficient choice is
	\begin{equation}
	h(x)\le\min\{h_0(x),h_1(x)\}.
	\end{equation}
This bounds the phase or relative variation by $1/2$. The height $h_0(x)$ controls where the estimate is valid, while $h_1(x)$ supplies a sufficient bound for positivity. Positive local lower bounds allow a Lipschitz minorant. Equation~\eqref{eq:vertical} and conjugation exclude nonreal preimages throughout the ribbon. Step~4 checks the endpoints.

	\paragraph*{Example.}
	For GIG, the phase estimate holds for every $y\ge0$, so the positivity bound alone determines the height. Set
	\begin{equation}\label{eq:gigpath}
	h(x)=h_1(x)=\frac{x^2}{b+2|a-1|x+bx^2},\qquad x>0.
	\end{equation}
The vertical segments remain in $\Rea z>0$, and for $0\le y\le h(x)$,
	\begin{equation}
	|\arg g(x+\ii y)|
	\le h(x)\left(\frac{|a-1|}{x}+\frac b2+\frac{b}{2x^2}\right)
	=\frac12<\frac\pi2.
	\end{equation}
Thus one contour excludes nonreal preimages for every $0<p<1$.

\subsection{Step 4: Verify}\label{sec:step-verify}
Verify absolute convergence of the upper integral and vanishing of the vertical closing integrals. Given an envelope $|g(x+\ii y)|\le E(x)$ for $|y|\le h(x)$, a sufficient test is
	\begin{equation}\label{eq:endpoint-test}
	\int_I|\kappa(x)\kappa'(x)|E(x)\dd x<\infty,\qquad
	h(x)E(x)\longrightarrow0,\qquad
	h(x)(|x|+h(x))E(x)\longrightarrow0
	\end{equation}
at both endpoints. The second condition forces the complex CDF to approach $0$ and $1$ uniformly on the closing segments, so the denominator tends to $-p$ and $1-p$. The first and third conditions control the upper integral and the closings.
	
If this test fails, the compact-coordinate construction of Theorem~\ref{thm:universal} remains available. Another option is a holomorphic observable $G$, real on $I$ with a known inverse there. Replacing $\kappa$ in the kernel by $G(\kappa)$ returns $G(\mathcal Q(p))$, subject to new convergence and closing checks. A bounded observable accommodates Student laws without a first moment.
	
	\paragraph*{Example.}
For GIG, $h(x)=O(x^2)$ at zero, $h(x)\to1/b$ at infinity, and $\kappa'$ is bounded. Uniformly along the vertical segments, the density is bounded by a constant times $x^{a-1}e^{-b/(2x)}$ near zero and $x^{a-1}e^{-bx/2}$ near infinity. The upper integrand therefore has orders
	\begin{equation}
	O(x^a e^{-b/(2x)})\quad(x\downarrow0),\qquad
	O(x^a e^{-bx/2})\quad(x\to\infty).
	\end{equation}
Both bounds are integrable. The CDF displacement vanishes, and the closing integrals are $O(x^{a+2}e^{-b/(2x)})$ and $O(x^a e^{-bx/2})$, respectively. Hence \eqref{eq:endpoint-test} holds for every $a\in\R$ and $b>0$.

\subsection{Step 5: Explicit Quantile}\label{sec:step-quantile}
	Apply \eqref{eq:intro} or \eqref{eq:quantile-density}, once its endpoint conditions hold, to obtain the working quantile $\mathcal Q$. For $u=T(x)$, the original quantile is $\widetilde{\mathcal Q}(p)=T(\mathcal Q(p))$ if $T$ is increasing and $\widetilde{\mathcal Q}(p)=T(\mathcal Q(1-p))$ if it is decreasing. Without a coordinate change, $\widetilde{\mathcal Q}=\mathcal Q$. This includes restoring location and scale. If an observable was used, first invert $G$ on the real interval. Applying $T$ inside the contour integral instead requires a holomorphic continuation and new endpoint checks.
		
	Under the conditions of \eqref{eq:quantile-density}, integration by parts gives \eqref{eq:intro} when the logarithmic integral converges and the imaginary part of $\kappa\Log[(\mathcal P(\kappa)-p)/(1-p)]$ vanishes at both endpoints. The compact construction satisfies these conditions. Otherwise, the imaginary boundary term must be retained.
	
	\paragraph*{Example.}
For the GIG law, the scale reconstruction is
	\begin{equation}\label{eq:gig-reconstruction}
	\widetilde{\mathcal Q}_{\mathrm{GIG}}(p;\sigma,a,b)
	=\sigma\mathcal Q_{\mathrm{GIG}}(p),
	\end{equation}
and the working quantile is given by
	\begin{equation}\label{eq:gigquantile}
	\mathcal Q_{\mathrm{GIG}}(p)
	=\frac{1}{\pi}\Ima\int_0^\infty
	\kappa'(x)\Log\left(
	\frac{\mathcal P_{\mathrm{GIG}}(\kappa(x))-p}{1-p}
	\right)\dd x.
	\end{equation}
	For $h$ in \eqref{eq:gigpath}, the contour \eqref{eq:contour} has derivative
	\begin{equation}
	\kappa'(x)=1+\ii\frac{2bx+2|a-1|x^2}{(b+2|a-1|x+bx^2)^2}.
	\end{equation}
The estimates in Step~4 and the tail bound $1-\mathcal P_{\mathrm{GIG}}(\kappa(x))=O(x^{a-1}e^{-bx/2})$ ensure absolute convergence and a vanishing boundary term. The integrand uses only forward functions, and the contour is independent of $p$. 

\section{Example Distributions}\label{sec:families}
Each family PDF $f$ below is standardized, with original density $f((u-\mu)/\sigma)\, /\, \sigma$, where $\sigma>0$ and $\mu=0$ for GB, GG, and GIG. The interval $I$ always denotes the support of the working density $g$, which may differ from that of $f$. Shape symbols are local to each family. The working quantile $\mathcal Q$ is obtained by the method in Section~\ref{sec:method}. Derivations are collected in Appendix~\ref{app:family-details}.

\begin{table}[H]
\centering
\caption{Working densities, contour heights, and quantile reconstruction.}
\label{tab:gb-specializations}
\small
\renewcommand{\arraystretch}{0.8}
\setlength{\extrarowheight}{0pt}
\setlength{\tabcolsep}{2pt}
\begin{tabular}{lccc}
\toprule
Distribution &  $g(x)$ &  $h(x)$ &  $\widetilde{\mathcal Q}(p)$ \\
\midrule
\textbf{GB}
& $\displaystyle\frac{x^{a-1}(1-x)^{b-1}}{\mathrm B(a,b)}$
& $\displaystyle\frac{x(1-x)}{1+|a-1|(1-x)+|b-1|x}$
& $\displaystyle\sigma\left[\frac{\mathcal Q(p)}{1-\rho\mathcal Q(p)}\right]^{1/c}$ \\
\textbf{GG}
& $\displaystyle\frac{x^{a-1}e^{-x}}{\Gamma(a)}$
& $\displaystyle\frac{x}{1+|a-1|+x}$
& $\sigma[\mathcal Q(p)]^{1/c}$ \\
\textbf{GIG}
& see Eq.~\eqref{eq:gigdensity}
& $\displaystyle\frac{x^2}{b+2|a-1|x+bx^2}$
& $\sigma\mathcal Q(p)$ \\
\textbf{GH}
& $g_{\mathrm{GH}}(x)$
& $h_{\mathrm{GH}}$
& $\mu+\sigma\mathcal Q(p)$ \\
\textbf{GEV}, $a\ne0$
& $e^{-x}$
& $\displaystyle\frac{x}{1+x}$
& $\displaystyle\mu+\frac{\sigma}{a}\left([\mathcal Q(1-p)]^{-a}-1\right)$ \\
\textbf{GEV}, $a=0$
& $e^{-x}$
& $\displaystyle\frac{x}{1+x}$
& $\mu-\sigma\log\mathcal Q(1-p)$ \\
Normal
& $\phi(x)$
& $\displaystyle\frac{1}{2(1+x^2)}$
& $\mu+\sigma\mathcal Q(p)$ \\
Student's t
& $t_a(x)$
& $\displaystyle\frac{\sqrt a}{2(a+1)}$
& $\mu+\sigma\mathcal Q(p)$ \\
Cauchy
& $\displaystyle\frac{1}{\pi(1+x^2)}$
& $1/4$
& $\mu+\sigma\mathcal Q(p)$ \\
Variance gamma
& $g_\pm(x)$
& $h_{\mathrm{VG}}(x)$
& $\mu+\sigma\mathcal Q(p)$ \\
Log-normal
& $\phi(x)$
& $\displaystyle\frac{1}{2(1+x^2)}$
& $\sigma\exp[a\mathcal Q(p)]$ \\
\bottomrule
\end{tabular}

\medskip
\singlespacing
\raggedright
Here $\phi$ is the standard normal density, and $t_a$ is the Student density with $a>0$ degrees of freedom. Student and Cauchy use the observable $G(z)=\arctan z$. For log-normal, $a>0$ is the log-scale standard deviation. For VG, $\mathcal Q$ denotes the full standardized quantile after the half-axis probability adjustment. The working densities, contour heights, and reconstruction details are given in Appendix~\ref{app:family-details}.
\end{table}

\begin{list}{\textbullet}{\setlength{\leftmargin}{1em}\setlength{\labelwidth}{0.5em}\setlength{\labelsep}{0.5em}\setlength{\itemsep}{0.35\baselineskip}\setlength{\parsep}{0pt}\setlength{\topsep}{0.25\baselineskip}\setlength{\partopsep}{0pt}}
\item \phantomsection\label{sec:gb}\textbf{Generalized beta (GB)}\\*
The working interval is $I=(0,1)$. The standardized family density is \citep{McDonaldXu1995}
	\begin{equation}
		f(x)=\frac{c\,x^{ca-1}[1-(1-\rho)x^c]^{b-1}}
		{\mathrm B(a,b)(1+\rho x^c)^{a+b}},
		\qquad x>0,\quad (1-\rho)x^c<1,
	\end{equation}
Here $a,b,c>0$ and $0\le\rho\le1$. The map $u=\sigma[x/(1-\rho x)]^{1/c}$ takes the beta working law to GB, so only $a,b$ enter the contour. The remaining parameters enter the reconstruction. Special cases include GB1 ($\rho=0$), GB2 ($\rho=1$), beta ($c=1,\rho=0$), beta-prime ($c=1,\rho=1$), uniform ($a=b=c=1,\rho=0$), and log-logistic ($a=b=1,\rho=1$). Fisher F with degrees of freedom $2a,2b$ uses $c=1$, $\rho=1$, and $\sigma=b/a$.

\begin{samepage}
\item \phantomsection\label{sec:gg}\textbf{Generalized gamma (GG)}\\*
The working interval is $I=(0,\infty)$, and the standardized family density is\par
\end{samepage}
	\begin{equation}
	f(x)=\frac{c}{\Gamma(a)}x^{ca-1}e^{-x^c},
	\qquad x>0,\quad a,c>0.
	\end{equation}
The map $u=\sigma x^{1/c}$ takes a unit-rate gamma law of shape $a$ to GG. Gamma has $c=1$, Weibull has $a=1$, and exponential has $a=c=1$. Chi-squared with $2a$ degrees of freedom uses $c=1$ and $\sigma=2$. Unit-scale half-normal, Rayleigh, and Maxwell use $c=2$, $\sigma=\sqrt2$, and $a=1/2,1,3/2$, respectively. GG is itself a rescaled limit of GB.

\item \phantomsection\label{sec:gig}\textbf{Generalized inverse Gaussian (GIG)}\\*
The working interval is $I=(0,\infty)$, with $g=f$ as in \eqref{eq:gigdensity}, $a\in\R$, and $b>0$. The map $u=\sigma x$ restores scale. Inverse Gaussian and reciprocal inverse Gaussian have $a=-1/2$ and $a=1/2$, respectively. For inverse Gaussian, $\sigma$ is the mean and $\sigma b$ the conventional shape. Section~\ref{sec:method} gives the construction and \eqref{eq:gigquantile}.

\item \phantomsection\label{sec:gh}\textbf{Generalized hyperbolic (GH)} \\*
The working interval is $I=\R$, and the standardized family density is \citep{BarndorffNielsen1977}
	\begin{equation}
	f(x)=
	\frac{(b^2-d^2)^{a/2}e^{dx}(1+x^2)^{(a-1/2)/2}}
	{\sqrt{2\pi}\,b^{a-1/2}K_a(\sqrt{b^2-d^2})}
	K_{a-1/2}\!\left(b\sqrt{1+x^2}\right),
	\qquad x\in\R,
	\end{equation}
Here $a\in\R$ is the order, $b>|d|$ controls tail decay, and $d$ controls skewness. I take $g_{\mathrm{GH}}=f$ and restore location and scale through $u=\mu+\sigma x$. Hyperbolic and normal inverse Gaussian have $a=1$ and $a=-1/2$, respectively.

\item \phantomsection\label{sec:gev}\textbf{Generalized extreme value (GEV)} \\*
The working interval is $I=(0,\infty)$. For $a\in\R$, the standardized family density is
\begin{equation}
f(x)=
\begin{cases}
(1+ax)^{-1/a-1}\exp[-(1+ax)^{-1/a}],&a\ne0,\quad 1+ax>0,\\
\exp[-x-e^{-x}],&a=0,\quad x\in\R.
\end{cases}
\end{equation}
The decreasing map $u=\mu+\sigma(x^{-a}-1)/a$ for $a\ne0$, or $u=\mu-\sigma\log x$ for $a=0$, takes a unit-rate exponential law to GEV. Reconstruction therefore uses $1-p$. Fr\'echet, Gumbel, and reverse Weibull correspond to $a>0$, $a=0$, and $a<0$. Substituting $\mathcal Q(1-p)=-\log p$ recovers their elementary quantile formulas.

\item \phantomsection\label{sec:limits}\textbf{Limiting cases.} Normal and log-normal use a standard normal working density on $I=\R$. For log-normal, $\mu=0$, $\sigma$ is the median, and $a>0$ is the log-scale standard deviation. Student's t uses its standardized density on $I=\R$ and the observable $G(z)=\arctan z$ to cover all degrees of freedom $a>0$. Cauchy is the case $a=1$. 

VG is treated through two analytic half-axis subdensities on $I=(0,\infty)$. The sign of $p-\mathcal P(0)$ selects the half-axis, and $|p-\mathcal P(0)|$ sets the target mass. The affine map then restores location and scale. Here $a>-1/2$ is the Bessel order and $d\in\R$ controls skewness.

\end{list}

\section{Example Application: Implied Volatility}\label{sec:applications}
Implied volatility is the standard quoting convention in option markets and provides a direct finance application of the method. In the Black--Scholes model, it is the volatility parameter $\sigma_{\mathrm{BS}}$ that reproduces an observed option price through the pricing formula \citep{BlackScholes1973}. Despite its widespread use, analytical inversion remains difficult. The connection to an inverse-Gaussian quantile established in \cite{Schadner2026Quantile} allows us to apply the GIG construction.
	
Let $\mathsf c=\mathsf{C}/(\mathsf{D}\mathsf{F})\in(0,1)$ be the call price $\mathsf{C}$ normalized by the discount factor $\mathsf{D}>0$ and forward price $\mathsf{F}>0$, and let $\bar{\mathsf c}$ denote its observed value. Write $\mathsf{k}=\log(\mathsf{K}/\mathsf{F})>0$ for moneyness, where $\mathsf{K}>0$ is the strike, $\mathsf{T}>0$ for maturity, and $v=\sigma_{\mathrm{BS}}\sqrt{\mathsf{T}}$ for total volatility. With $\Phi$ the standard normal CDF, the Black--Scholes call price formula can be written as
	\begin{equation}\label{eq:c}
		\mathsf c(\mathsf{k},v)=\Phi\!\left(-\frac{\mathsf{k}}{v}+\frac{v}{2}\right)-e^{\mathsf{k}}\Phi\!\left(-\frac{\mathsf{k}}{v}-\frac{v}{2}\right)=1-\mathcal P_{\mathrm{IG}}\!\left(\frac{4}{v^2};\frac{2}{\mathsf{k}},1\right).
	\end{equation}
Here $\mathcal P_{\mathrm{IG}}$ is parametrized by its mean and shape. In the GIG convention used here, the corresponding parameters are $(\sigma,a,b)=(2/\mathsf{k},-1/2,\mathsf{k}/2)$. Consequently, $\sigma_{\mathrm{BS}}^2\mathsf{T}$ is the $\bar{\mathsf c}$-quantile of the GIG law with parameters $(\sigma,a,b)=(2\mathsf{k},1/2,\mathsf{k}/2)$, obtained by the reciprocal transformation $u\mapsto4/u$. Scaling the GIG construction \eqref{eq:gigpath} by $2\mathsf{k}$ gives
	\begin{equation}\label{eq:blackpath}
		\kappa_{\mathsf{k}}(x)=x+\ii\frac{x^2}{\mathsf{k}^2+x+x^2/4},\qquad x>0.
	\end{equation}
After the square-root change from variance to total volatility, the contour can be inserted directly into \eqref{eq:intro} or \eqref{eq:quantile-density} to obtain an explicit implied-volatility formula. Instead, I prefer a variant with integration over the finite interval $(0,1)$. Define $\nu_{\mathsf{k}}(t)=\sqrt{\kappa_{\mathsf{k}}(t/(1-t))}$ for $0<t<1$, combining the substitution $x=t/(1-t)$ with the map to total volatility. By the chain rule, $\nu_{\mathsf{k}}'(t)=\kappa_{\mathsf{k}}'(t/(1-t))/(2(1-t)^2\nu_{\mathsf{k}}(t))$, so
	\begin{equation}\label{eq:blackintegral}
		\sigma_{\mathrm{BS}}=\frac{1}{\pi\sqrt{\mathsf{T}}}\Ima\int_0^1
		\nu_{\mathsf{k}}'(t)
		\Log\!\left(\frac{\mathsf c(\mathsf{k},\nu_{\mathsf{k}}(t))-\bar{\mathsf c}}{1-\bar{\mathsf c}}\right)\dd t.
	\end{equation}
Hence, implied volatility has an explicit integral representation involving only elementary functions and the error function. The logarithm's argument compares the model price along the contour with the observed price. It vanishes at the sought real total volatility, by \eqref{eq:c}, while the contour and its reflection enclose this zero without passing through it.

The integral converges absolutely for $\mathsf{k}>0$ and $0<\bar{\mathsf c}<1$ with principal branches for the complex powers and logarithm. Put--call parity gives $\mathsf c(\mathsf{k},v)=1-e^{\mathsf{k}}+e^{\mathsf{k}}\mathsf c(-\mathsf{k},v)$, reducing $\mathsf{k}<0$ to the positive branch. At $\mathsf{k}=0$, the price reduces to a normal-CDF expression \citep{Schadner2026Quantile}.
	
A finance reader will further ask for an explicit formula for Bachelier (normal) implied volatility. Appendix~\ref{app:normal-details} derives the corresponding representation.

\section{Conclusion}\label{sec:conclusion}
The construction separates contour selection from the probability level, allowing one verified contour to represent an entire quantile function. Existence holds for strictly positive real-analytic densities after a known coordinate change, while elementary heights are available for the families studied here. The Black--Scholes application illustrates the resulting formulas in finance. Quantifying quadrature error, particularly near extreme probability levels, remains a useful next step.

\newpage
\bibliographystyle{elsarticle-harv}
\bibliography{literature}

\newpage
\appendix
\counterwithin{equation}{section}
\counterwithin{table}{section}
\counterwithin{figure}{section}
\section*{Online Appendix}
\section{Distribution-specific contour derivations}\label{app:family-details}
This appendix derives the working densities and contour heights used in Section~\ref{sec:families}. Each construction establishes positivity along vertical segments and then checks the support endpoints. Powers are continued from their positive real values, and CDFs are holomorphic primitives normalized to agree with their real counterparts \citepapp{DLMF}. The GIG derivation is already given in Section~\ref{sec:method}.

\subsection{Generalized beta (GB)}\label{app:gb}
For the GB family, the map $u=\sigma[x/(1-\rho x)]^{1/c}$ has derivative $(\sigma/c)x^{1/c-1}(1-\rho x)^{-1/c-1}>0$. Substitution in the original density gives the beta working law. Its two powers are holomorphic on $0<\Rea z<1$, with branches fixed by the real density. For $y\ge0$,
\begin{equation}
|\arg g(x+\ii y)|
\le |a-1|\arctan\frac yx+|b-1|\arctan\frac y{1-x}
\le y\left(\frac{|a-1|}{x}+\frac{|b-1|}{1-x}\right).
\end{equation}
To keep the phase below one while forcing the height to vanish at both endpoints, add $1/[x(1-x)]$ to the coefficient of $y$ and take the reciprocal. This gives
\begin{equation}\label{eq:gbheight}
h(x)=\frac{x(1-x)}{1+|a-1|(1-x)+|b-1|x},\qquad 0<x<1.
\end{equation}
The phase satisfies $|\arg g|\le1<\pi/2$ throughout the ribbon. Since $h\le\min\{x,1-x\}$, the modulus of $g$ there is bounded by a constant times $x^{a-1}(1-x)^{b-1}$. Also $\kappa'$ is bounded. The working-quantile kernel has orders $O(x^a)$ and $O((1-x)^{b-1})$. The CDF displacement vanishes at both endpoints, and the vertical closing contributions have orders $O(x^{a+1})$ and $O((1-x)^b)$. These estimates establish the inversion conditions for every $a,b>0$, including densities that diverge at an endpoint. Applying the increasing map recovers the GB quantile without a moment assumption on the transformed law. Appendix~\ref{app:gb2} gives a direct GB2 integral with endpoint regularization.

\subsection{Generalized gamma and extreme-value laws}\label{app:gg}\label{app:gev}
For GG, the map $u=\sigma x^{1/c}$ and its Jacobian $(\sigma/c)x^{1/c-1}$ reduce the density to the unit-rate gamma law. On the right half-plane,
\begin{equation}
\arg g(x+\ii y)=(a-1)\arctan(y/x)-y,\qquad
|\arg g(x+\ii y)|\le y\left(\frac{|a-1|}{x}+1\right).
\end{equation}
Adding $1/x$ to the coefficient of $y$ keeps the phase below one and makes the height vanish at zero. Taking the reciprocal gives
\begin{equation}\label{eq:ggheight}
h(x)=\frac{x}{1+|a-1|+x},\qquad x>0.
\end{equation}
The phase satisfies $|\arg g|\le1<\pi/2$ throughout the ribbon. Since $h\le\min\{x,1\}$ and $\kappa'$ is bounded, the density envelope is a constant times $x^{a-1}e^{-x}$. The working-quantile kernel has orders $O(x^a)$ at zero and $O(x^ae^{-x})$ at infinity. The CDF displacement vanishes at both endpoints, and the closing contributions are $O(x^{a+1})$ and $O(x^ae^{-x})$. Thus Step~4 holds for every $a>0$, and the increasing map gives the reconstruction in Section~\ref{sec:gg}.
Setting $a=1$ simplifies the height to $x/(1+x)$, giving the construction for the exponential working law.

\paragraph*{GEV.}
For the GEV family with $a\ne0$, the map $u=\mu+\sigma(x^{-a}-1)/a$ has derivative $-\sigma x^{-a-1}<0$. At $a=0$, it becomes $u=\mu-\sigma\log x$, with derivative $-\sigma/x<0$. The original CDF at the transformed value is $e^{-x}$, so the working CDF is $1-e^{-x}$. The gamma construction at shape one therefore applies with $h(x)=x/(1+x)$. Since the map is decreasing, reconstruction uses $\mathcal Q(1-p)=-\log p$ and gives the elementary GEV quantile.

\subsection{Generalized hyperbolic (GH)}\label{app:gh-density}\label{app:gh}
For the GH working law $g=f$, with $a\in\R$ and $b>|d|$, use the constant height
\begin{equation}\label{eq:ghheight}
h(x)=\min\left\{\frac12,\,
\frac{e^{-b(1-\sqrt3/2)}(\sqrt3/2)^{3/2+|a-1/2|-a}}
{2|d|+4b/\sqrt3+\tfrac83(3/2+|a-1/2|-a)}
\right\},\qquad x\in\R.
\end{equation}
\begin{proposition}[GH contour]\label{prop:gh}
Let $a\in\R$ and $b>|d|$. With the height \eqref{eq:ghheight}, the GH working density satisfies $|g(x+\ii y)-g(x)|\le g(x)/2$ for $x\in\R$ and $0\le y\le h$. Consequently $\Rea g>0$ on the ribbon, and \eqref{eq:quantile-density} holds for every $0<p<1$.
\end{proposition}
The relative variation bound also transfers the real exponential tail bounds to the contour.

\begin{proof}
To bound the variation uniformly in $x$, set
\begin{equation}
D=1+|{{a}-\tfrac12}|-{{a}-\tfrac12},\quad
A_*=\exp[{b}(1-\sqrt3/2)](2/\sqrt3)^D,\quad
M_*=A_*(|{d}|+2{b}/\sqrt3+4D/3).
\end{equation}
The height in \eqref{eq:ghheight} is $\min\{1/2,1/(2M_*)\}$.

The strip $|\Ima z|<1$ has $\Rea[1+z^2]>0$, so the principal square root and power are consistent. Put $r(z)=\sqrt{1+z^2}$. The Bessel integral gives a representation locally uniform on this strip,
\begin{equation}
H(r):=r^{{a}-\tfrac12} K_{{a}-\tfrac12}({b} r)
=\frac{{b}^{{a}-\tfrac12}}{2}\int_0^\infty
t^{{{a}-\tfrac12}-1}\exp[-({b}^2t+r^2/t)/2]\dd t.
\end{equation}
For $z=u+\ii y$, $|y|\le1/2$, let $r=\sqrt{u^2+1}$ and $s=\sqrt{u^2+3/4}$. Differentiating under the integral and taking absolute values gives
\begin{equation}\label{eq:ghderivative}
\frac{|g'(u+\ii y)|}{g(u)}
\le\frac{H(s)}{H(r)}
\left[|{d}|+\frac{|u|+1/2}{s}L(s)\right],
\quad
L(s)={b}\frac{K_{{{a}-\tfrac12}-1}({b} s)}{K_{{a}-\tfrac12}({b} s)}.
\end{equation}
Indeed the differentiated integrand has multiplier ${d}-(u+\ii y)/t$, and the real part of $r(z)^2$ is at least $s^2$. No complex Bessel denominator is involved.

The positive integral gives $L(v)=-H'(v)/H(v)>0$. By Lemma~\ref{lem:bessel}, $L(v)\le{b}+D/v$. Consequently
\begin{equation}
\frac{H(s)}{H(r)}
=\exp\left(\int_s^rL(v)\dd v\right)
\le e^{{b}(r-s)}(r/s)^D\le A_*.
\end{equation}
Here $r-s\le1-\sqrt3/2$ and $r/s\le2/\sqrt3$. Also $(|u|+1/2)/s\le2/\sqrt3$ and $s\ge\sqrt3/2$. Equation \eqref{eq:ghderivative} is thus bounded by $M_*$. Vertical integration gives $|g(x+\ii y)-g(x)|\le g(x)/2$ for $0\le y\le h$.

For fixed parameters, the real tails have order $|u|^{a-1}e^{-{b}|u|+{d} u}$. The bound $|g(x+\ii y)|\le3g(x)/2$ transfers integrability to the ribbon. Also $|\mathcal P(x+\ii y)-\mathcal P(x)|\le3hg(x)/2\to0$. The observable $G(z)=z$ is integrable against this envelope, and the vertical closing integrals vanish. Proposition~\ref{prop:residue} therefore applies.
\end{proof}
The affine map in Section~\ref{sec:gh} restores location and scale. The following subsections construct separate contours for the boundary and limiting laws.

\subsection{Normal and log-normal limiting laws}\label{app:normal-lognormal}
Normal and log-normal use the same standard normal working law. The contour follows directly from a phase bound.

For the standard normal density $g(x)=f(x)=e^{-x^2/2}/\sqrt{2\pi}$, choose the height
\begin{equation}
h(x)=\frac{1}{2(1+x^2)},\qquad x\in\R.
\end{equation}
The continuous phase satisfies $|\arg g(x+\ii y)|=|xy|\le |x|/[2(1+x^2)]\le1/4$ for $0\le y\le h(x)$. Also $|g(x+\ii y)|/g(x)=e^{y^2/2}\le e^{1/8}$. Gaussian tails therefore give absolute convergence and vanishing closing integrals in \eqref{eq:quantile-density}. The affine map restores the normal location and scale.

For the log-normal limit of GG \citepapp{Bell1988}, take $\sigma,a>0$ and $\mu=0$. Its standardized density is
\begin{equation}
f(x)=\frac{1}{a x\sqrt{2\pi}}\exp\!\left[-\frac{(\log x)^2}{2a^2}\right],
\qquad x>0.
\end{equation}
The map $u=\sigma e^{ax}$ reduces inversion to the standard normal working law and its contour above. Here $a$ is the log-scale standard deviation and $\sigma$ the median. The reconstruction is given in Table~\ref{tab:gb-specializations}.

\subsection{Variance gamma and splitting at nonanalytic points}\label{sec:vg}\label{app:vg-formula}\label{app:split}\label{app:vg}
The VG density may fail to be analytic at zero. I therefore split the law there and invert each half-axis subdensity. The same reduction applies whenever a density is analytic on separate intervals. Let $\widetilde{\mathcal P}$ be the original CDF, and let $(\ell,r)$ be an interval on which its density is strictly positive and real analytic. The known endpoint probabilities determine the mass $m=\widetilde{\mathcal P}(r)-\widetilde{\mathcal P}(\ell)>0$. For a target between these endpoint probabilities, set
\begin{equation}
\mathcal P(x)=\frac{\widetilde{\mathcal P}(x)-\widetilde{\mathcal P}(\ell)}{m},
\qquad
\widehat p=\frac{p-\widetilde{\mathcal P}(\ell)}{m},
\qquad \ell<x<r.
\end{equation}
The working density is the original density divided by $m$, and $\widetilde{\mathcal Q}(p)=\mathcal Q(\widehat p)$. The mass cancels from the logarithmic derivative, so one can equivalently use the unnormalized sub-CDF $\widetilde{\mathcal P}(x)-\widetilde{\mathcal P}(\ell)$ and target $p-\widetilde{\mathcal P}(\ell)$. If $p$ equals the CDF value at a dividing point, that point is the quantile. Each analytic interval requires its own contour and endpoint estimates. The VG construction below uses this sub-CDF form.

Take ${a}>-1/2$ and ${d}\in\R$, representing the Bessel order and dimensionless skewness. The standardized density is \citepapp{Gaunt2023,FischerGauntSarantsev2025}
\begin{equation}
f(x)=\frac{e^{{d} x}|x|^{a} K_{a}({\sqrt{1+{d}^2}}|x|)}
{\sqrt{\pi}(2{\sqrt{1+{d}^2}})^{a}\Gamma({a}+1/2)}.
\end{equation}
At zero, the leading orders are $|x|^{2{a}}$ for $-1/2<{a}<0$, $\log(1/|x|)$ for ${a}=0$, and a finite positive limit for ${a}>0$. I treat the two half-axes separately. For $\varepsilon\in\{-1,1\}$ and $\Rea z>0$, define
\begin{equation}\label{eq:vg-subdensities}
\begin{aligned}
g_\varepsilon(z)&=
\frac{e^{\varepsilon{d} z}z^{a} K_{a}({\sqrt{1+{d}^2}} z)}
{\sqrt{\pi}(2{\sqrt{1+{d}^2}})^{a}\Gamma({a}+1/2)},\\
\mathcal P_\varepsilon(z)&=\int_0^zg_\varepsilon(u)\dd u.
\end{aligned}
\end{equation}
Writing $\mathcal P$ for the full standardized VG CDF, these sub-CDFs satisfy $\mathcal P_+(x)=\mathcal P(x)-\mathcal P(0)$ and $\mathcal P_-(x)=\mathcal P(0)-\mathcal P(-x)$. The boundary singularity is integrable.

\begin{theorem}[VG quantiles on two half-axes]\label{thm:vg}
Use
\begin{equation}\label{eq:vgcenter}
\begin{aligned}
\kappa(x)&=x+\ii\frac{x}
{8\,2^{|{a}|}[1+2|{a}|+({\sqrt{1+{d}^2}}+|{d}|)x]},
\\
p_0&=I_{({\sqrt{1+{d}^2}}-{d})/(2{\sqrt{1+{d}^2}})}
       ({a}+\tfrac12,{a}+\tfrac12).
\end{aligned}
\end{equation}
For $p\ne p_0$, put $\varepsilon=\sgn(p-p_0)$ and $\Delta p=|p-p_0|$. Then
\begin{equation}\label{eq:vgquantile}
\widetilde{\mathcal Q}(p)=\mu-\frac{\varepsilon\sigma}{\pi}\Ima
\int_0^\infty \kappa(x)
\frac{g_\varepsilon(\kappa(x))\kappa'(x)}
{\mathcal P_\varepsilon(\kappa(x))-\Delta p}\dd x.
\end{equation}
The full complex integral converges absolutely, and $\widetilde{\mathcal Q}(p_0)=\mu$.
\end{theorem}
\begin{proof}
To simplify the estimates, write $D=2^{|{a}|}$, $A=1+2|{a}|$, and $B={\sqrt{1+{d}^2}}+|{d}|$.

For $z=x+\ii y$, $0\le y\le x/2$, the positive Bessel integral gives $|K_{a}({\sqrt{1+{d}^2}} z)|\le K_{a}({\sqrt{1+{d}^2}} x)$ and $|K_{a}'({\sqrt{1+{d}^2}} z)|\le-K_{a}'({\sqrt{1+{d}^2}} x)$. Moreover $|z^{a}|/x^{a}\le D$. Differentiating $g_\varepsilon$, using $|z|^{-1}\le x^{-1}$, and applying Lemma~\ref{lem:bessel}, we obtain
\begin{equation}
\frac{|g_\varepsilon'(x+\ii y)|}{g_\varepsilon(x)}
\le D\left(|{d}|+\frac{|{a}|}{x}
             -{\sqrt{1+{d}^2}}\frac{K_{a}'({\sqrt{1+{d}^2}} x)}{K_{a}({\sqrt{1+{d}^2}} x)}\right)
\le D(B+A/x).
\end{equation}
The chosen height satisfies $h\le x/8$ and $hD(B+A/x)=1/8$, so $|g_\varepsilon(x+\ii y)-g_\varepsilon(x)|\le g_\varepsilon(x)/8$ throughout its ribbon. In particular $\Rea g_\varepsilon>0$ and $\Ima \mathcal P_\varepsilon>0$ above the axis.

At zero the identity-observable kernel is $O(xg_\varepsilon(x))$, which is integrable for every ${a}>-1/2$. The closing contribution is $O(x^2g_\varepsilon(x))$ and vanishes. At infinity, $h$ is bounded and $g_\varepsilon(x)=O(x^{{a}-1/2}e^{-({\sqrt{1+{d}^2}}-\varepsilon{d})x})$. Both tail rates are positive. For $0<\Delta p<\mathcal P_\varepsilon(\infty)$ the denominator has nonzero endpoint limits, proving the other convergence and closing conditions.

To determine which half-axis contains the quantile, represent the standardized law as the difference $U-V$ of independent gamma variables with common shape $a+1/2$ and rates ${\sqrt{1+{d}^2}}-{d},{\sqrt{1+{d}^2}}+{d}$. Multiplying each variable by its rate gives independent, identically distributed unit-rate gamma variables. The event $U<V$ corresponds to a beta ratio below $({\sqrt{1+{d}^2}}-{d})/(2{\sqrt{1+{d}^2}})$, proving $\mathcal P(0)=p_0$. The unique positive root of $\mathcal P_\varepsilon(r)=\Delta p$ gives $\widetilde{\mathcal Q}(p)=\mu+\varepsilon\sigma r$. The residue identity proves \eqref{eq:vgquantile}.
\end{proof}

\subsection{Student's t and Cauchy boundary laws}\label{app:student}
For Student's t, use the standardized density directly. For $a>0$,
\begin{equation}\label{eq:student-density}
g(x)=f(x)=t_a(x)=
\frac{\Gamma((a+1)/2)}{\sqrt{a\pi}\,\Gamma(a/2)}
\left(1+\frac{x^2}{a}\right)^{-(a+1)/2},
\qquad x\in\R.
\end{equation}
To cover all degrees of freedom, including laws without a first absolute moment, use the bounded observable
\begin{equation}\label{eq:theight}
h(x)=\frac{\sqrt a}{2(a+1)},\qquad G(z)=\arctan z.
\end{equation}
The bounds $h(x)<\sqrt a$ and $h(x)\le1/4$ place the density's singularities and those of $G$ outside the ribbon. Moreover
\begin{equation}
|\arg g(x+\ii y)|
\le\frac{({a}+1)y}{2\sqrt{{a}-y^2}}
\le\frac1{4\sqrt{1-[2({a}+1)]^{-2}}}<\frac\pi2.
\end{equation}
For $0\le y\le h$, the argument estimate follows from $\arctan v\le v$ and $2|x|/(a+x^2-y^2)\le1/\sqrt{a-y^2}$. Also $h^2/a\le1/4$, so $\Rea(1+z^2/a)\ge(3/4)(1+x^2/a)$ and $|g(z)|\le(4/3)^{(a+1)/2}g(x)$. Since $G$ is bounded on this ribbon, the kernel has tails $O(|x|^{-a-1})$ and the closing integrals vanish. The height is constant, so $\kappa'=1$, and the observable formula gives
\begin{equation}\label{eq:t}
\widetilde{\mathcal Q}(p)=\mu+\sigma\tan\left[-\frac1\pi\Ima\int_{\R}
\arctan(\kappa(x))\frac{g(\kappa(x))}{\mathcal P(\kappa(x))-p}\dd x\right].
\end{equation}
The formula covers all $a>0$, including $a\le1$, where the identity-observable integral need not be absolutely convergent. The Cauchy law corresponds to $a=1$. Squaring a Student variable gives a Fisher law with degrees of freedom $(1,a)$, linking this family to GB. Recovering the signed quantile from that law requires a probability adjustment using symmetry. Appendix~\ref{app:sanity-checks} evaluates the bounded-observable integral at one degree of freedom and the identity-observable integral at two.

\subsection{Bessel estimate used for GH and VG}\label{app:bessel}
\begin{lemma}\label{lem:bessel}
For every real $a$ and $t>0$,
\begin{equation}
0<-\frac{K_a'(t)}{K_a(t)}
\le1+\frac{|a|+1}{t}.
\end{equation}
\end{lemma}
\begin{proof}
Use the classical positive integral \citepapp[Eq.~10.32.9]{DLMF}
\begin{equation}
K_a(t)=\int_0^\infty e^{-t\cosh v}\cosh(|a|v)\dd v.
\end{equation}
Normalize its integrand to a probability measure, and put $m=\E\cosh v=-K_a'/K_a$. Integration of the derivative of $e^{-t\cosh v}\sinh v\cosh(|a|v)$, whose endpoint terms vanish, gives
\begin{equation}
t\,\E\sinh^2v
=\E[\cosh v+|a|\sinh v\tanh(|a|v)]
\le(1+|a|)m.
\end{equation}
Jensen's inequality yields $t(m^2-1)\le(1+|a|)m$, which implies the stated upper bound.
\end{proof}

\section{Proofs and extensions of the inversion method}\label{app:proofs}
The residue identity isolates the quantile once a contour has been constructed. I first prove that identity and a sufficient endpoint test, then establish existence and the quantitative disk construction. The remaining subsections derive the logarithmic form and the extensions used in the distribution and option-pricing examples.

\subsection{Residue identity and observables}\label{app:residue}\label{sec:residue}
For a positively oriented simple closed contour $\mathcal C$ and a function $H$ holomorphic on and inside it, with no zeros on $\mathcal C$, the classical root-moment identity is
\begin{equation}\label{eq:classicalroots}
\frac{1}{2\pi\ii}\oint_{\mathcal C} z^N\frac{H'(z)}{H(z)}\dd z
=\sum_j z_j^N,\qquad N=0,1,2,\ldots,
\end{equation}
where the zeros inside $\mathcal C$ are counted with multiplicity \citepapp[Eq.~(1.4), p.~545]{DelvesLyness1967Appendix}.

\begin{proposition}[Classical residue identity with endpoint conditions]
\label{prop:residue}
Let $g=\mathcal P'$ be a strictly positive normalized density on $I$, and let $h>0$ be locally Lipschitz. Suppose $\mathcal P$ and $g$ extend holomorphically to a neighborhood of $\mathcal R_h$, agree with their real values on $I$, and satisfy
\begin{equation}
\Ima \mathcal P(x+\ii y)>0\quad(0<y\le h(x)).
\end{equation}
Let $G$ be holomorphic there and real on $I$. For fixed $0<p<1$ assume the integral in \eqref{eq:observable} converges absolutely and the integrals of $G(z)g(z)/(\mathcal P(z)-p)$ over the vertical closing segments vanish as the truncated interval tends to $I$. Then
\begin{equation}\label{eq:observable}
G(\mathcal Q(p))=-\frac1\pi\Ima\int_I
G(\kappa(x))\kappa'(x)\frac{g(\kappa(x))}{\mathcal P(\kappa(x))-p}\dd x.
\end{equation}
\end{proposition}
\begin{proof}
Conjugation excludes nonreal zeros of $\mathcal P-p$ below the axis as well as above it. There is exactly one real zero $\mathcal Q(p)$, and it is simple because $g(\mathcal Q(p))>0$. The residue of $Gg/(\mathcal P-p)$ is $G(\mathcal Q(p))$. Traverse the upper graph left to right and its reflection right to left, closing vertically. This gives a clockwise contour. The two graph integrals sum to $2\ii$ times the imaginary part of the upper integral. The residue theorem and the endpoint hypotheses give \eqref{eq:observable}.
\end{proof}

\subsection{A sufficient endpoint test}\label{app:endpoints}
\begin{proposition}[Endpoint test]\label{prop:endpoints}
Under the holomorphy and sign assumptions of Section~\ref{sec:method}, suppose $|g(x+\ii y)|\le E(x)$ throughout the ribbon and the three conditions in \eqref{eq:endpoint-test} hold. Then the inversion integral is absolutely convergent and the vertical closing contributions vanish for every fixed $0<p<1$.
\end{proposition}
\begin{proof}
Vertical integration gives
\begin{equation}
|\mathcal P(x+\ii y)-\mathcal P(x)|\le h(x)E(x)\longrightarrow0
\end{equation}
uniformly for $|y|\le h(x)$ at each endpoint. Since the real CDF tends to $0$ and $1$, the denominator is bounded away from zero near the corresponding ends. On each compact subinterval, the sign condition excludes zeros on the upper graph, so the denominator has a positive minimum modulus. The first condition in \eqref{eq:endpoint-test} therefore bounds the absolute integral. Each vertical closing contribution is bounded by a constant times $h(x)(|x|+h(x))E(x)$ and tends to zero.
\end{proof}
For an observable $G$, replace $|\kappa(x)|$ in the integrability test by $|G(\kappa(x))|$ and replace $|x|+h(x)$ in the closing test by $\sup_{|y|\le h(x)}|G(x+\ii y)|$. The CDF-displacement condition remains $hE\to0$.

\subsection{Compact coordinates and proof of the existence theorem}\label{app:universal}
Under the assumptions of Theorem~\ref{thm:universal}, let $\widetilde{\mathcal P}$ and $\widetilde{\mathcal Q}$ be the original CDF and quantile. Choose an increasing real-analytic bijection $\psi:(0,1)\to I$ with $\psi'>0$, and put
\begin{equation}
\mathcal P=\widetilde{\mathcal P}\circ\psi,\qquad g=\mathcal P',\qquad \mathcal Q(p)=\psi^{-1}(\widetilde{\mathcal Q}(p)).
\end{equation}
It suffices to construct a positive Lipschitz height on $(0,1)$, independent of $p$, for which
\begin{equation}\label{eq:universal}
\mathcal Q(p)=-\frac1\pi\Ima\int_0^1
\kappa(x)\kappa'(x)\frac{g(\kappa(x))}{\mathcal P(\kappa(x))-p}\dd x,\qquad
\widetilde{\mathcal Q}(p)=\psi(\mathcal Q(p)).
\end{equation}
Useful choices are $\psi(x)=x/(1-x)$ on $(0,\infty)$, $\psi(x)=\tan(\pi(x-\tfrac12))$ on $\R$, and affine maps for finite intervals.
\begin{proof}[Proof of Theorem~\ref{thm:universal}]
Local holomorphic continuations can be joined along the real interval and restricted to a simply connected, symmetric thin neighborhood. Put $d(x)=\min\{x,1-x\}$. Local continuity and positivity give admissible radii with positive local lower bounds. Taking a continuous positive minorant yields a radius $\rho$ such that the vertical segments lie in that neighborhood and
\begin{equation}
\rho(x)\le\frac{d(x)}{4(1+g(x))},\qquad
|g(x+\ii y)-g(x)|\le\tfrac12g(x)\quad(|y|\le\rho(x)).
\end{equation}
Define
\begin{equation}\label{eq:minorant}
h(x)=\frac12\inf_{0<s<1}\{\rho(s)+|x-s|\}.
\end{equation}
This is $1/2$-Lipschitz and $h\le\rho/2$. For fixed interior $x$, split the infimum into a compact neighborhood of $x$ and its complement. Positivity of $\rho$ on the first and positive distance on the second show $h(x)>0$.

Almost everywhere $|\kappa'|\le\sqrt5/2$, and on the ribbon
\begin{equation}
|g(x+\ii y)|\le\tfrac32g(x),\qquad
|\mathcal P(x+\ii y)-\mathcal P(x)|\le\tfrac32h(x)g(x)\le\tfrac3{16}d(x).
\end{equation}
Consequently $\mathcal P$ tends uniformly to $0$ and $1$ on the respective end segments. Its imaginary part is positive on the upper graph by \eqref{eq:vertical}. For fixed $p$, the denominator in \eqref{eq:universal} is bounded away from zero near the endpoints and on every compact middle part. Boundedness of $\kappa,\kappa'$ and $\int_0^1g=1$ prove absolute convergence. The closing integrals are bounded by constants times $h(x)g(x)$ and vanish. Proposition~\ref{prop:residue} with $G(z)=z$ gives \eqref{eq:universal}. The construction uses no moment of the original law.
\end{proof}

\subsection{A quantitative construction from complex disks}\label{app:disk}
If a derived upper bound $\widehat h:I\to(0,\infty)$ has positive local lower bounds, the formula
\begin{equation}
h(x)=\inf_{s\in I}\{\widehat h(s)+\tfrac12|x-s|\}
\end{equation}
produces a positive $1/2$-Lipschitz function with $h\le\widehat h$. Positivity follows by splitting the infimum into a compact neighborhood of $x$, where $\widehat h$ has a positive lower bound, and its complement, where $|x-s|$ is bounded below. Thus any positive local height bound can be replaced by a Lipschitz height suitable for Step~3.

\begin{proposition}[Disk majorants]\label{prop:disk}
Put $d(x)=\min\{x,1-x\}$. In the compact coordinates of Appendix~\ref{app:universal}, suppose $g$ is holomorphic near each closed disk $\overline D(x,R(x))$ and
\begin{equation}
\sup_{|z-x|\le R(x)}|g(z)|\le B(x),\qquad B(x)\ge g(x)>0.
\end{equation}
Assume $R$ is locally bounded below away from zero and $B$ is locally bounded above. Put
\begin{equation}
b(x)=\min\left\{\frac{d(x)}{8(1+g(x))},
                    \frac{g(x)R(x)}{8B(x)}\right\},\qquad
h(x)=\inf_{0<s<1}\{b(s)+\tfrac12|x-s|\}.
\end{equation}
Then this $h$ is positive, $1/2$-Lipschitz, and admissible in \eqref{eq:universal}.
\end{proposition}
\begin{proof}
The minorant is positive by the local bounds and $h\le b$. Cauchy's coefficient estimate gives, for $0\le y\le h(x)\le R(x)/8$,
\begin{equation}
|g(x+\ii y)-g(x)|\le B(x)\frac{y}{R(x)-y}
\le\frac17g(x).
\end{equation}
Thus $\Rea g\ge6g/7$ and $|g|\le8g/7$ on the ribbon. Also $|\mathcal P(x+\ii y)-\mathcal P(x)|\le d(x)/7$. The endpoint argument in Appendix~\ref{app:universal} applies.
\end{proof}
Applying this prescription requires explicit disk bounds derived from the density. The proposition does not infer them from finitely many real-axis evaluations.

\subsection{Logarithmic representation and endpoint terms}\label{app:log}
For a general admissible contour, put $L_p(x)=\Log[(\mathcal P(\kappa(x))-p)/(1-p)]$, with argument in $(0,\pi)$. Integration by parts on a truncated interval and passage to the endpoint limits give
\begin{equation}
\mathcal Q(p)=\frac1\pi\Ima\int_I\kappa'(x)L_p(x)\dd x
-\frac1\pi\Ima[\kappa(x)L_p(x)]_{\partial I},
\end{equation}
whenever these limits exist. The bracket denotes the right endpoint value minus the left endpoint value. Thus the logarithmic representation requires a separate boundary check before applying \eqref{eq:intro}.

For the compact construction of Theorem~\ref{thm:universal}, this condition holds and gives
\begin{equation}\label{eq:log}
\mathcal Q(p)=\frac1\pi\Ima\int_0^1
\kappa'(x)\Log\left(\frac{\mathcal P(\kappa(x))-p}{1-p}\right)\dd x.
\end{equation}
The bounds $h(x)\le d(x)/8$ in Appendices~\ref{app:universal} and~\ref{app:disk} imply $\kappa\to0$ and $\kappa\to1$ at the respective endpoints. The logarithm tends to $\log[p/(1-p)]+\ii\pi$ at zero and to zero at one. It is also bounded on every compact subinterval because its argument stays in the upper half-plane. The integral is absolutely convergent because $\kappa'$ is bounded and the logarithm has finite endpoint limits.

\subsection{Decreasing analytic functions}\label{app:decreasing}
Let $H$ be real analytic on $I$ with $H'<0$, and let $\zeta\in H(I)$. Suppose $H$ extends holomorphically to a neighborhood of the ribbon and $\Rea H'(x+\ii y)<0$ for $0\le y\le h(x)$. Then
\begin{equation}
\Ima H(x+\ii y)=\int_0^y\Rea H'(x+\ii t)\dd t<0
\end{equation}
for $y>0$. Conjugation excludes nonreal zeros below the axis, while $H'<0$ gives one simple real zero. The proof of Proposition~\ref{prop:residue} therefore yields
\begin{equation}
H^{-1}(\zeta)=-\frac1\pi\Ima\int_I
\kappa(x)\kappa'(x)\frac{H'(\kappa(x))}{H(\kappa(x))-\zeta}\dd x,
\end{equation}
provided this integral is absolutely convergent and the vertical closing integrals vanish. The derivative $H'$ enters with its original sign. The hypotheses concern the complete kernel, so neither normalization of $H$ nor integrability of $|H'|$ alone is required. The same argument permits a holomorphic observable in place of $\kappa$.

\subsection{Endpoint regularization for a direct GB2 integral}\label{app:gb2}
In the beta working coordinate, one may multiply the residue kernel by a holomorphic factor $R_p(z)$, real on $I$, satisfying $R_p(\mathcal Q(p))=1$. Reality on $I$ ensures conjugate symmetry, so the two arcs still combine into an imaginary part. The factor preserves the desired residue provided no additional singularities are introduced and the modified endpoint integrals converge. Factors constructed from $\mathcal P(z)$, for example $[(1-\mathcal P(z))/(1-p)]^m$, suppress a troublesome endpoint without using the unknown quantile.
For the direct GB2 observable, choose a nonnegative integer $m$ with $(m+1){b}>1/{c}$, for example $m=\lfloor1/({c} {b})\rfloor$. Using the beta working CDF, density and contour from Appendix~\ref{app:gb} gives
\begin{equation}\label{eq:gb2}
\widetilde{\mathcal Q}(p)=-\frac{\sigma}{\pi}\Ima\int_0^1
\left(\frac{\kappa(x)}{1-\kappa(x)}\right)^{1/{c}}
\frac{g(\kappa(x))\kappa'(x)}{\mathcal P(\kappa(x))-p}
\left(\frac{1-\mathcal P(\kappa(x))}{1-p}\right)^m\dd x.
\end{equation}
The upper-integrand endpoint orders are $x^{{a}+1/{c}-1}$ and $(1-x)^{(m+1){b}-1-1/{c}}$, both integrable. The corresponding closing integrals vanish. The regularizing factor depends on $p$, while the contour remains fixed.

\subsection{Power observables and logarithmic integrals for GIG}\label{app:gig-powers}
For the standardized GIG law in Section~\ref{sec:step-density}, the factors $e^{-b/(2x)}$ and $e^{-bx/2}$ control every real power at zero and infinity. Proposition~\ref{prop:residue} therefore gives, for any real $N$,
\begin{equation}\label{eq:gig-power-density}
[\mathcal Q_{\mathrm{GIG}}(p)]^N
=-\frac1\pi\Ima\int_0^\infty
\kappa(x)^N\kappa'(x)
\frac{g(\kappa(x))}{\mathcal P_{\mathrm{GIG}}(\kappa(x))-p}\dd x.
\end{equation}
The powers are continued from the positive real axis. The integral converges absolutely and its closing integrals vanish.

For $N>0$, integration by parts also gives
\begin{equation}\label{eq:gig-power-log}
[\mathcal Q_{\mathrm{GIG}}(p)]^N
=\frac{N}{\pi}\Ima\int_0^\infty
\kappa(x)^{N-1}\kappa'(x)
\Log\!\left(\frac{\mathcal P_{\mathrm{GIG}}(\kappa(x))-p}{1-p}\right)\dd x.
\end{equation}
At zero the logarithm is bounded and $\kappa^N\to0$. At infinity, $1-\mathcal P_{\mathrm{GIG}}(\kappa(x))=O(x^{a-1}e^{-bx/2})$, so the logarithm has the same order and $\kappa^N$ times the logarithm tends to zero. The logarithmic integrand is $O(x^{N-1})$ at zero and exponentially small at infinity. Thus the boundary term vanishes and the integral converges absolutely.

Taking $N=1$ gives \eqref{eq:gigquantile}. For $N=1/2$, the transformations in Section~\ref{sec:applications} give the implied-volatility formula \eqref{eq:blackintegral}.

\section{Exact sanity checks}\label{app:sanity-checks}
I evaluate the prescribed contour integrals exactly for Student laws with one and two degrees of freedom. Recovering their elementary quantiles checks the normalization, contour orientation, and logarithm branches. Throughout, $0<p<1$ and $\sigma>0$.

\subsection{One degree of freedom (Cauchy)}
For $a=1$, the standardized density, CDF, and prescribed contour are
\begin{equation}
g(z)=\frac{1}{\pi(1+z^2)},\qquad
\mathcal P(z)=\frac12+\frac{\arctan z}{\pi},
\qquad
\kappa_1(x)=x+\frac{\ii}{4},\quad x\in\R.
\end{equation}
Use the bounded observable $G(z)=\arctan z$ from \eqref{eq:t}, since the identity-observable integral is not absolutely convergent. Define
\begin{equation}
J_1(p)=\int_{\R}
\arctan(\kappa_1(x))\,\kappa_1'(x)
\frac{g(\kappa_1(x))}
{\mathcal P(\kappa_1(x))-p}\dd x.
\end{equation}
The bounded-observable representation gives
\begin{equation}
\widetilde{\mathcal Q}(p)=\mu+\sigma\tan\left[-\frac1\pi\Ima J_1(p)\right].
\end{equation}
Here $\kappa_1'=1$, and the integrand is $O(|x|^{-2})$ at both ends.

Set $w=\mathcal P(\kappa_1(x))$. Then $\dd w=g(\kappa_1(x))\kappa_1'(x)\dd x$ and $\arctan(\kappa_1(x))=\pi(w-1/2)$. The image contour $\gamma_1$ runs from $0$ to $1$ in the upper half-plane. Consequently
\begin{equation}
\begin{aligned}
J_1(p)
&=\pi\int_{\gamma_1}\frac{w-\tfrac12}{w-p}\dd w\\
&=\pi\left[w+\left(p-\frac12\right)\Log(w-p)\right]_{0}^{1}.
\end{aligned}
\end{equation}
Along the upper-half-plane image contour, the logarithm has endpoint values $\log p+\ii\pi$ at $w=0$ and $\log(1-p)$ at $w=1$. Thus
\begin{equation}\label{eq:sanity-cauchy-integral}
J_1(p)=\pi\left\{
1+\left(p-\frac12\right)
\left[\log\frac{1-p}{p}-\ii\pi\right]
\right\}.
\end{equation}
It follows that $-\Ima J_1(p)/\pi=\pi(p-1/2)$, which lies in the real range of $\arctan$. Hence
\begin{equation}\label{eq:sanity-cauchy-quantile}
\widetilde{\mathcal Q}(p)=\mu+\sigma\tan\!\left[\pi\left(p-\frac12\right)\right].
\end{equation}

\subsection{Two degrees of freedom}
For $a=2$, the standardized density, CDF, and contour are
\begin{equation}
g(z)=\frac{1}{(z^2+2)^{3/2}},\qquad
\mathcal P(z)=\frac12+\frac{z}{2\sqrt{z^2+2}},
\qquad
\kappa_2(x)=x+\ii\frac{\sqrt2}{6}.
\end{equation}
Choose the square root positive on the real axis and holomorphic on $|\Ima z|<\sqrt2$. This law has a finite first absolute moment, so the identity observable is admissible. Put
\begin{equation}
J_2(p)=\int_{\R}
\kappa_2(x)\kappa_2'(x)
\frac{g(\kappa_2(x))}
{\mathcal P(\kappa_2(x))-p}\dd x.
\end{equation}
Then $\widetilde{\mathcal Q}(p)=\mu-\sigma\Ima J_2(p)/\pi$. The integrand is $O(|x|^{-2})$ and the vertical closing contributions vanish.

Set
\begin{equation}
w=\frac{z+\sqrt{z^2+2}}{\sqrt2},
\qquad
z=\frac{w-w^{-1}}{\sqrt2}.
\end{equation}
This substitution makes both the CDF and the integration kernel rational in $w$,
\begin{equation}
\mathcal P(z)=\frac{w^2}{1+w^2},
\qquad
g(z)\dd z=\frac{2w}{(1+w^2)^2}\dd w.
\end{equation}
The contour $\kappa_2$ maps to a curve $\gamma_2$ in the first quadrant, running from $0$ to infinity. Indeed, on this contour $\Rea\sqrt{z^2+2}>|\Rea z|$ and $|\Ima\sqrt{z^2+2}|<\Ima z$, so both components of $w$ are positive. Substitution and partial fractions give
\begin{equation}
\begin{aligned}
J_2(p)
&=\sqrt2\int_{\gamma_2}
\frac{w^2-1}{(1+w^2)((1-p)w^2-p)}\dd w\\
&=\sqrt2\int_{\gamma_2}
\left[\frac{2}{1+w^2}+\frac{2p-1}{(1-p)w^2-p}\right]\dd w.
\end{aligned}
\end{equation}
With $\alpha=\sqrt{p/(1-p)}>0$, an antiderivative is
\begin{equation}
2\sqrt2\,\arctan w
+\frac{2p-1}{\sqrt{2p(1-p)}}
\Log\left(\frac{w-\alpha}{w+\alpha}\right).
\end{equation}
Along $\gamma_2$, the ratio inside the logarithm has positive imaginary part. Its logarithm tends to $\ii\pi$ at zero and to $0$ at infinity, while $\arctan w$ tends to $0$ and $\pi/2$. Hence
\begin{equation}\label{eq:sanity-t2-integral}
J_2(p)=\sqrt2\,\pi-\ii\pi\frac{2p-1}{\sqrt{2p(1-p)}}.
\end{equation}
Taking the imaginary part yields
\begin{equation}\label{eq:sanity-t2-quantile}
\widetilde{\mathcal Q}(p)=\mu+\sigma\frac{2p-1}{\sqrt{2p(1-p)}}.
\end{equation}
Substitution of the standardized quantile into $\mathcal P$ gives $p$, completing the identity-observable check.

\section{Numerical Verifications}\label{app:numerical}
I compare direct contour-integral evaluations with independently computed reference values. The first experiment covers the working laws in Section~\ref{sec:families}. The second tests the Black--Scholes implied-volatility formula in Section~\ref{sec:applications}. The reported discrepancies measure agreement on these test sets and do not constitute uniform quadrature error bounds.

\subsection{Quantiles of the example distributions}\label{app:numerical-quantiles}
I evaluate the residue form \eqref{eq:quantile-density} with the heights of Section~\ref{sec:families} for twelve working laws at eight probability levels and compare the result with reference quantiles. The laws are beta with $(a,b)=(0.3,0.6)$ and $(2.5,3)$, gamma with $a=0.4$ and $a=1$, the latter being the exponential working law of GEV, GIG with $(a,b)=(-1/2,1)$, GH with $(a,b,d)=(1,1.5,0.5)$ and $(-1/2,2,1)$, the standard normal, Student's t with $a=1/2$ and $a=5$, and VG with $(a,d)=(-0.3,0.15)$ and $(1.5,0.2)$. Student's t uses the observable in \eqref{eq:t}, and VG the half-axis representation of Theorem~\ref{thm:vg}. The reported errors concern the working quantiles, before applying the reconstruction maps.

Each integral uses composite 20-point Gauss--Legendre quadrature in 24-digit arithmetic with mpmath. The cells lie in a logistic, logarithmic, or linear coordinate and are no wider than the contour height in that coordinate, except in the far Student tails beyond all tested quantiles. Each law uses one mesh for all eight levels. Where no closed-form complex CDF is available (GIG, GH, VG), $\mathcal P(\kappa)$ is accumulated along the contour with the matching Gauss--Legendre integration matrix. Reference quantiles solve $\mathcal P(x)=p$ by Newton's method in 30-digit arithmetic.

Reference CDFs use closed forms for the beta, gamma, normal, and Student laws and double-exponential quadrature of the real density for GIG, GH, and VG. For VG with $a<0$, the substitution $y=xu^{1/(2a+1)}$ removes the singularity at zero, and the mass below zero is integrated rather than taken from $p_0$. Newton's method starts from double-precision quantiles computed with SciPy. Accumulation of $\mathcal P(\kappa)$ starts where the neglected mass is below $e^{-60}$, at $x=b/130$ for GIG and at the left truncation point for GH. For VG, it starts after a straight segment from $0$ to $\kappa(10^{-14})$, evaluated with the same substitution. Each domain is truncated where the neglected integrand is of order $e^{-60}$. As controls, the accumulated CDFs reach their total masses, $1$ for GIG and GH and the half-axis masses implied by $p_0$ for VG, to about $10^{-24}$, and the numerically integrated reference densities integrate to one within $10^{-30}$. The implementation is available at \url{https://github.com/wol-fi/quantile}.

\begin{figure}[h]
\centering
\includegraphics[width=\linewidth]{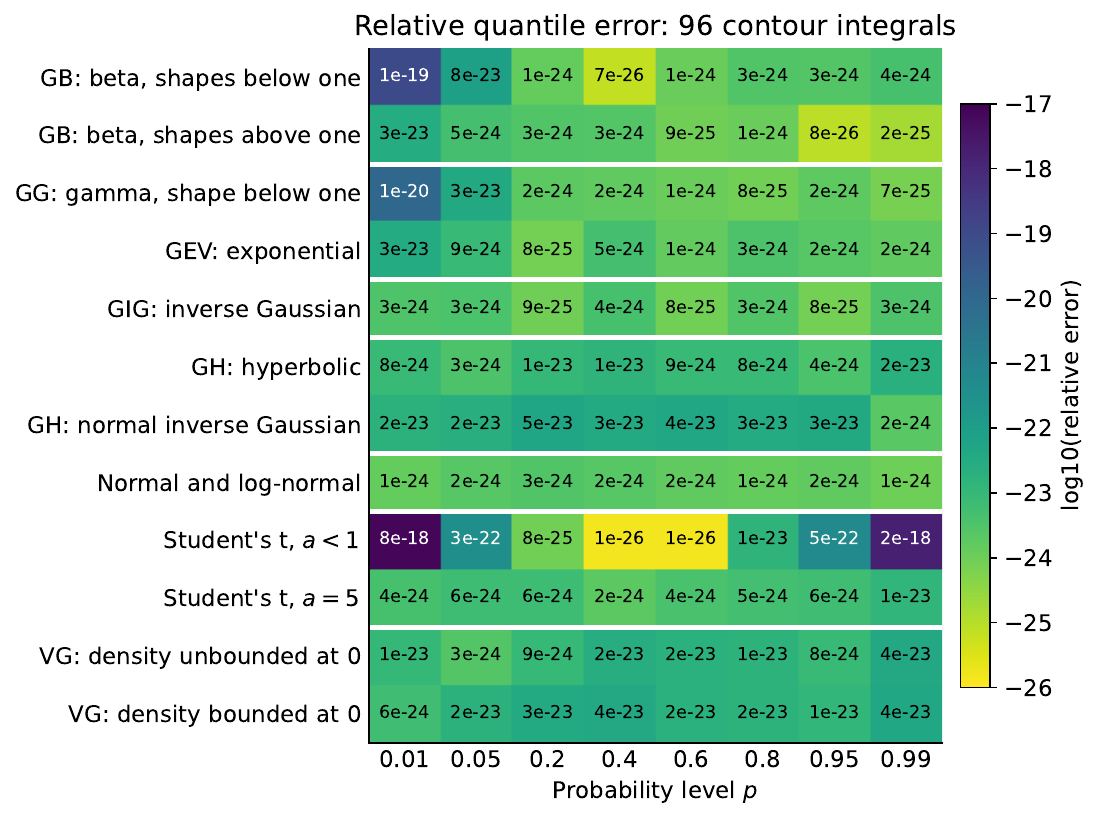}
\caption{Relative errors of the contour-integral quantiles in Appendix~\ref{app:numerical-quantiles}. Rows are working laws, grouped by family, and columns are probability levels. Cells give the error to one significant digit. Colors show a logarithmic error scale, distinct from that in Figure~\ref{fig:bs-verification}.}
\label{fig:quantile-verification}
\end{figure}

Figure~\ref{fig:quantile-verification} shows the relative errors. All 96 are below $10^{-17}$, and most lie between $10^{-25}$ and $10^{-22}$. Cancellation and the observable inversion make some cases more sensitive to finite working precision. At the smallest quantiles, beta and gamma at $p=0.01$, contributions of order one cancel to a value of order $10^{-7}$ or $10^{-6}$. For Student's t with $a=1/2$, the quantiles at $p=0.01$ and $0.99$ are about $\pm10^3$, and the tangent reconstruction amplifies relative errors in the arctangent observable by a factor of the same order.

\subsection{Black--Scholes implied volatility}\label{app:numerical-bs}
\begin{figure}[htbp]
\centering
\includegraphics[width=\linewidth]{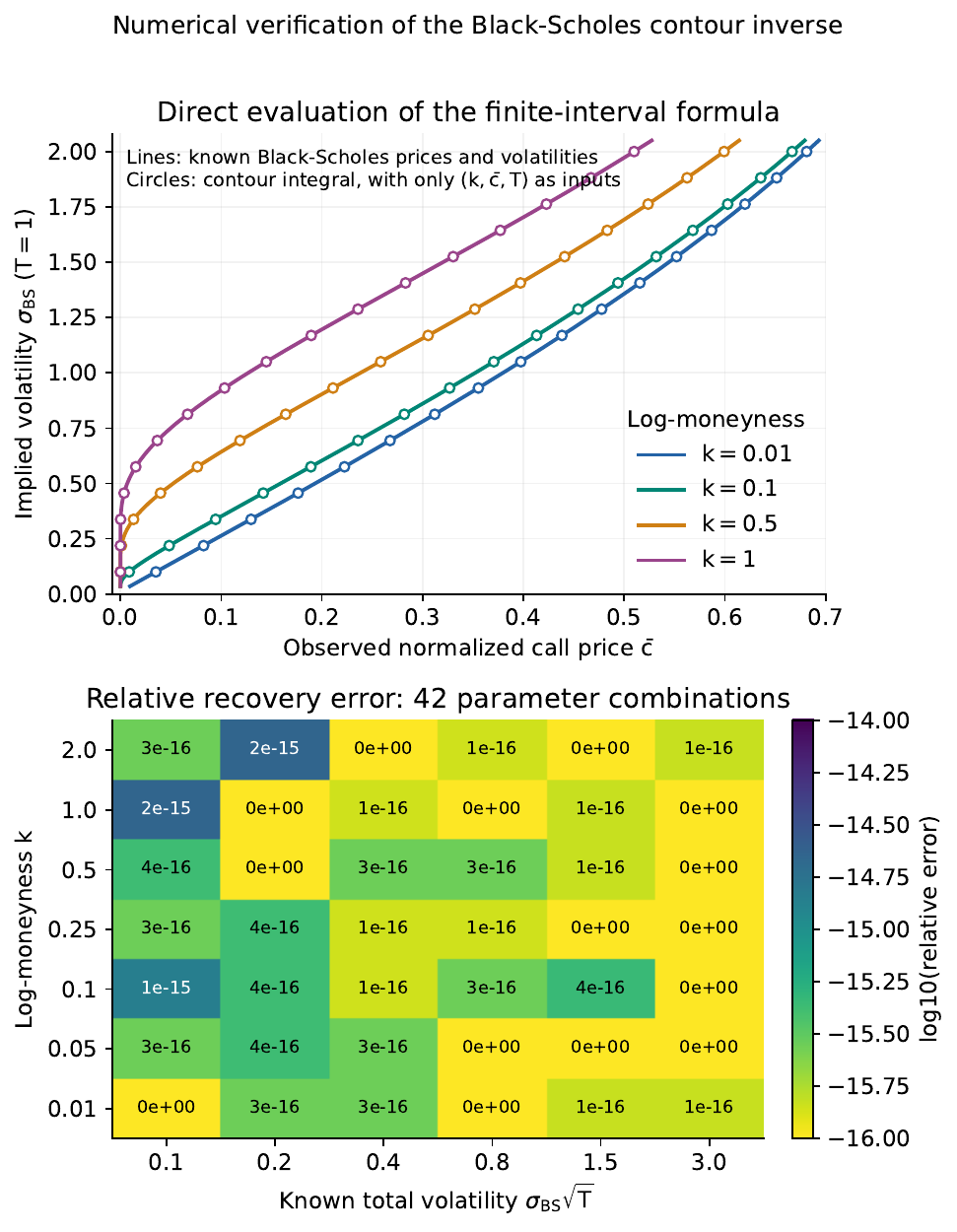}
\caption{Numerical verification of \eqref{eq:blackintegral}. In the upper panel, lines give reference Black--Scholes price--volatility pairs and circles give direct contour-integral evaluations. The lower panel shows relative volatility errors for the 42-case grid. Zero entries denote agreement after floating-point rounding, and the color scale is floored at $10^{-16}$.}
\label{fig:bs-verification}
\end{figure}
I test \eqref{eq:blackintegral} by generating forward-normalized call prices from known volatilities and recovering them through the contour integral. The reference prices are evaluated from the Black--Scholes formula at 100 decimal digits. The integral evaluator receives only $(\mathsf{k},\bar{\mathsf c},\mathsf{T})$. Its initial subdivision is independent of the reference volatility, and recovery uses quadrature without root finding.

The test grid comprises all 42 combinations of $\mathsf{k}\in\{0.01,0.05,0.1,0.25,0.5,1,2\}$ and $\sigma_{\mathrm{BS}}\sqrt{\mathsf{T}}\in\{0.1,0.2,0.4,0.8,1.5,3\}$, with maturities $\mathsf{T}\in\{0.25,1,4\}$ assigned cyclically across the cases. The resulting call prices range from $3.72\times10^{-91}$ to $0.866$. Double-precision adaptive quadrature in SciPy uses absolute and relative tolerances of $2\times10^{-12}$. The substitution $t=s^2$ removes the integrable square-root singularity at zero, and neither endpoint is truncated. Algebraically equivalent evaluations of the logarithm avoid cancellation when the IG CDF is close to zero or one, while preserving the specified branch. If the imaginary part underflows near the left endpoint, the implementation must retain the upper-half-plane limit of the logarithm, with argument $+\pi$.

Across the grid, the largest absolute and relative volatility errors are $1.33\times10^{-15}$ and $2.22\times10^{-15}$, respectively. Repricing the recovered volatilities gives a largest absolute call-price error of $2.22\times10^{-16}$. Figure~\ref{fig:bs-verification} also shows 68 integral evaluations at unit maturity, all with relative errors at most $2.22\times10^{-15}$. The quadrature routine issued no warnings.

Independent checks in mpmath evaluate the IG CDF in its conventional mean--shape parametrization and use Gauss--Legendre quadrature at 35 and 50 decimal digits. For $(\mathsf{k},\sigma_{\mathrm{BS}}\sqrt{\mathsf{T}})=(0.05,0.4),(0.5,0.8),(1,1.5)$, the largest observed relative discrepancy at 50 digits is $1.67\times10^{-51}$. These computations support the normalization, maturity scaling, coordinate changes, and logarithm branch in \eqref{eq:blackintegral} over the tested range.

\FloatBarrier

\section{Bachelier implied volatility}\label{app:volatility}\label{app:normal-details}
	The financial variables $\mathsf{C}$, $\mathsf{D}$, $\mathsf{F}$, $\mathsf{K}$, and $\mathsf{T}$ are as in Section~\ref{sec:applications}, except that the Bachelier model also allows $\mathsf{F},\mathsf{K}\le0$. Put $\mathsf{m}=|\mathsf{F}-\mathsf{K}|>0$ and define the observed normalized time value $\bar{\mathrm V}=[\mathsf{C}/\mathsf{D}-(\mathsf{F}-\mathsf{K})_+]/\mathsf{m}>0$. Write $\phi$ and $\Phi$ for the standard normal density and CDF, and $v_{\mathrm N}=\sigma_{\mathrm N}\sqrt{\mathsf{T}}/\mathsf{m}$ for the normalized total volatility. The implied-volatility equation is $\mathrm V(v_{\mathrm N})=\bar{\mathrm V}$, where $\mathrm V$ is the normalized Bachelier time-value function
	\begin{equation}\label{eq:normalB}
	\mathrm V(v)=v\,\phi(1/v)-\Phi(-1/v),\qquad v>0.
	\end{equation}
	Equivalently, $\mathrm V(v)=\Gamma(-1/2,1/(2v^2))/(4\sqrt\pi)$, the incomplete-gamma form of the Bachelier time value \citepapp[Proposition~1]{Grunspan2011Normal} that is also used in normal-volatility inversion \citepapp{Jackel2017Normal}.
	Differentiation gives $\mathrm V'(v)=\phi(1/v)>0$, and $\mathrm V$ increases from $0$ to $\infty$. Hence $\sigma_{\mathrm N}=(\mathsf{m}/\sqrt{\mathsf{T}})\,\mathrm V^{-1}(\bar{\mathrm V})$. The function $\mathrm V$ is not a distribution function, but it continues holomorphically to $\C\setminus\{0\}$ because $\phi$ and $\Phi$ are entire. With the contour
	\begin{equation}\label{eq:normalpath}
	\kappa_{\mathrm N}(x)=x+\ii\frac{x^3}{2+x^3},\qquad
	\kappa_{\mathrm N}'(x)=1+\ii\frac{6x^2}{(2+x^3)^2},\qquad x>0,
	\end{equation}
	normal implied volatility has the representation
	\begin{equation}\label{eq:normalintegral}
		\sigma_{\mathrm N}=\frac{\mathsf{m}}{\pi\sqrt{\mathsf{T}}}\left(1-\int_0^\infty
		\Ima\left[\frac{\kappa_{\mathrm N}(x)\kappa_{\mathrm N}'(x)\,\phi(1/\kappa_{\mathrm N}(x))}
		{\mathrm V(\kappa_{\mathrm N}(x))-\bar{\mathrm V}}\right]\dd x\right).
	\end{equation}

The logarithmic variant is more compact on a finite interval. Write $\nu(t)=\kappa_{\mathrm N}(t/(1-t))$ for $0<t<1$. Then
	\begin{equation}\label{eq:normal-log-integral}
	\sigma_{\mathrm N}
	=\frac{\mathsf{m}}{\pi\sqrt{\mathsf{T}}}
	\int_0^1\Ima\!\left[
	\nu'(t)\Log\!\left(
	\frac{\mathrm V(\nu(t))-\bar{\mathrm V}}{\mathrm V(\nu(t))+\bar{\mathrm V}}
	\right)\right]\dd t.
	\end{equation}

	To verify \eqref{eq:normalintegral}, let $v_{\mathrm N}=\mathrm V^{-1}(\bar{\mathrm V})$, so that $\sigma_{\mathrm N}=(\mathsf{m}/\sqrt{\mathsf{T}})\,v_{\mathrm N}$. For $z=x+\ii y$, the continuous phase of $\mathrm V'(z)=\phi(1/z)$ is
	\begin{equation}
		\arg \mathrm V'(x+\ii y)=\frac{xy}{(x^2+y^2)^2}\le\frac{y}{x^3}.
	\end{equation}
	For $0\le y\le x^3/(2+x^3)$ it is therefore at most $1/(2+x^3)\le1/2$, so $\Rea \mathrm V'>0$ and $\Ima \mathrm V(x+\ii y)=\int_0^y\Rea \mathrm V'(x+\ii t)\dd t>0$ for $y>0$. Together with conjugation, this leaves $v_{\mathrm N}$ as the only zero of $\mathrm V-\bar{\mathrm V}$ in the ribbon, and it is simple. The residue of $z\mathrm V'(z)/(\mathrm V(z)-\bar{\mathrm V})$ at $v_{\mathrm N}$ is $v_{\mathrm N}$.
	
	At zero, $\Rea(1/z^2)\ge1/(2x^2)$ on the ribbon for small $x$, so $\phi(1/z)$ is $O(e^{-1/(4x^2)})$, while $\mathrm V(z)\to0$ and the denominator tends to $-\bar{\mathrm V}$. The integrand and the closing contribution at the left endpoint therefore vanish. At infinity, uniformly on the ribbon,
	\begin{equation}
		\mathrm V(z)=\frac{z}{\sqrt{2\pi}}-\frac12+O(z^{-1}),\qquad
		\frac{z\mathrm V'(z)}{\mathrm V(z)-\bar{\mathrm V}}=1+\frac{\sqrt{2\pi}(\bar{\mathrm V}+\tfrac12)}{z}+O(z^{-2}).
	\end{equation}
	The imaginary part of the integrand is thus $h'(x)-\sqrt{2\pi}(\bar{\mathrm V}+\tfrac12)h(x)/|\kappa_{\mathrm N}(x)|^2+O(x^{-2})$ with $h(x)=x^3/(2+x^3)$, which is $O(x^{-2})$ and integrable. As in Appendix~\ref{app:decreasing}, the function inverted need not be normalized. Unlike there, its vertical closing segment at $x=R$ does not vanish. The integrand tends to one uniformly, so the segment contributes $-2\ii h(R)+o(1)$ when the contour is traversed clockwise as in the proof of Proposition~\ref{prop:residue}. The residue theorem gives
	\begin{equation}
		-2\pi\ii\,v_{\mathrm N}=2\ii\int_0^R\Ima\left[\frac{\kappa_{\mathrm N}\kappa_{\mathrm N}'\,\phi(1/\kappa_{\mathrm N})}{\mathrm V(\kappa_{\mathrm N})-\bar{\mathrm V}}\right]\dd x-2\ii h(R)+o(1),
	\end{equation}
	and $h(R)\to1$ yields \eqref{eq:normalintegral}. The contour is independent of the target and the option parameters.

To obtain \eqref{eq:normal-log-integral}, use the same contour with the logarithm $L_{\mathrm N}(z)=\Log[(\mathrm V(z)-\bar{\mathrm V})/(\mathrm V(z)+\bar{\mathrm V})]$. On the upper ribbon, $\Ima \mathrm V(z)>0$, so the ratio has positive imaginary part and the logarithm takes its branch with argument in $(0,\pi)$. The factor $\mathrm V(z)+\bar{\mathrm V}$ has no zero in the ribbon or on its real base. Differentiating $L_{\mathrm N}$ subtracts the zero-free kernel $\mathrm V'(z)/(\mathrm V(z)+\bar{\mathrm V})$ from the kernel in \eqref{eq:normalintegral}; the residue remains $v_{\mathrm N}$. At zero, $L_{\mathrm N}(\kappa_{\mathrm N}(x))\to\ii\pi$ and $\kappa_{\mathrm N}(x)\to0$. At infinity, $L_{\mathrm N}(z)=-2\sqrt{2\pi}\,\bar{\mathrm V}/z+O(z^{-2})$, uniformly on the closing segments. Thus $\kappa_{\mathrm N}L_{\mathrm N}$ has a real limit, and its imaginary boundary term vanishes. Integration by parts gives $v_{\mathrm N}=\pi^{-1}\int_0^\infty\Ima[\kappa_{\mathrm N}'(x)L_{\mathrm N}(\kappa_{\mathrm N}(x))]\dd x$. The imaginary integrand is $O(x^{-2})$ at infinity, while its limit at zero is $\pi$. Substitution $x=t/(1-t)$ yields \eqref{eq:normal-log-integral}.
	
	At the money, $\sigma_{\mathrm N}=\sqrt{2\pi/\mathsf{T}}\,\mathsf{C}/\mathsf{D}$. Zero time value gives zero implied volatility by continuity. Every positive time value has a unique finite normal implied volatility.

\newpage
\FloatBarrier
\bibliographystyleapp{elsarticle-harv}
\bibliographyapp{literature}

\end{document}